\documentclass[11pt,a4paper]{article}
\pdfoutput=1

\usepackage{bm}
\usepackage{amsmath,amssymb}
\usepackage{cite}

\usepackage{graphicx}
\usepackage{color}
\usepackage{upgreek}

\usepackage{hyperref} 
\hypersetup{colorlinks=true, citecolor=blue, filecolor=black, linkcolor=blue, urlcolor=blue, pdfpagemode=UseNone}

\numberwithin{figure}{section}
\numberwithin{equation}{section}

\newcommand{\be}{\begin{equation}}
\newcommand{\ee}{\end{equation}}
\newcommand{\bea}{\begin{eqnarray}}
\newcommand{\eea}{\end{eqnarray}}
\def\beal#1\eeal{\begin{align}#1\end{align}}   
\def\besp#1\eesp{\begin{multline}#1\end{multline}} 

\newcommand{\TRM}[1]{#1}
\usepackage[normalem]{ulem}

\newcommand{\cL}{\mathcal{L}}

\newcommand{\ph}{\varphi}
\newcommand{\vp}{\varphi}

\newcommand{\vpi}{\uppi}

\newcommand{\Lz}{ {\Lambda_0} }

\newcommand{\p}{\mathrm{p}}
\newcommand{\dd}[2]{\delta_{\!\phantom{(} #1}^{\!(#2)}\!(\ph)}

\newcommand{\Sh}{\mathcal{S}}
\newcommand{\I}{\mathcal{I}}
\newcommand{\fV}{\mathcal{V}}

\newcommand{\M}{\mathcal{M}}

\newcommand{\h}{h }

\newcommand{\Latt}{\underline{\Lambda}}
\newcommand{\erfc}{\text{erfc}}

\newcommand\ie{\textit{i.e.}\ }
\newcommand\eg{\textit{e.g.}\ }

\newcommand{\half}{\tfrac{1}{2}}

\begin{document}

\begin{titlepage}

\begin{center}
{\huge \bf Renormalization group properties of the conformal mode of a torus}

\end{center}
\vskip1cm


\begin{center}
{\bf Matthew P. Kellett and Tim R. Morris}
\end{center}

\begin{center}
{\it STAG Research Centre \& Department of Physics and Astronomy,\\  University of Southampton,
Highfield, Southampton, SO17 1BJ, U.K.}\\
\vspace*{0.3cm}
{\tt M.P.Kellett@soton.ac.uk, T.R.Morris@soton.ac.uk}
\end{center}

\abstract{The Wilsonian renormalization group properties of the conformal factor of the metric are profoundly altered by the fact that it has a wrong-sign kinetic term. If couplings are chosen so that the quantum field theory exists on $\mathbb{R}^4$, it fails to exist on manifolds below a certain size, if a certain universal shape function turns negative. We demonstrate that this is triggered by inhomogeneity in the cases of  $\mathbb{T}^4$ and $\mathbb{T}^3\times\mathbb{R}$, including twisted versions. Varying the moduli, we uncover a rich phenomenology.}

\end{titlepage}

\tableofcontents

\newpage

\section{Introduction}
\label{sec:Intro}

In a recent paper \cite{Morris:2018mhd} it was shown that the conformal factor field, $\ph$, parametrising the overall `size' of the metric, has profound Wilsonian renormalization group (RG) properties, which hold out the prospect of constructing a perturbatively renormalizable quantum gravity. 

In order to study such RG properties one has to work with a Euclidean signature metric. In an $\mathbb{R}^4$ spacetime, after imposing a quantization condition on the bare action (reviewed later), the Gaussian fixed point supports infinite towers of relevant operators which span a Hilbert space of possible interactions. Even at the linearised level (\ie justified by vanishingly small couplings), complete flows exist to the infrared (IR) only for appropriate choices of couplings. Such solutions are characterised by a physical scale $\Lambda_\p\ge0$, beyond which the conformal factor of the metric is exponentially suppressed. 

The claim was made that on other spacetimes, 
of length scale $L$, the RG flow ends prematurely (and thus the quantum field theory itself ceases to exist) once a certain universal measure of inhomogeneity exceeds $O(1)+2\pi L^2\Lambda^2_\p$. 
If such a feature were to survive in a full theory of quantum gravity, it would clearly be important phenomenologically, the minimum size of the universe being  
thus tied to the degree of inhomogeneity. For example, it would explain why the initial conditions for inflation had to be sufficiently smooth. Since it ties the minimum size of the universe to the degree of inhomogeneity, and large amplitude inhomogeneities have appeared only recently in the history of the universe,  it could also explain the infamous ``Why now?'' problem, namely that the energy density of matter (including dark matter)  is now similar in magnitude to the apparent energy density of dark energy deduced from the current acceleration of the universe.  Finally for example, it implies the elimination of singularities, and thus ``cosmic censorship'' and somehow a softening of the causal structure of black holes.

The failure mechanism discovered in ref. \cite{Morris:2018mhd},
is a consequence of a certain universal dimensionless shape function $\Sh$ turning negative. The conclusion that this is triggered by universes with too much inhomogeneity, is based however on the study of a number of simple examples. In ref. \cite{Morris:2018mhd}, only the untwisted four-dimensional hyper-torus, $\mathbb{T}^4$, was presented. Here we deepen and generalise that work in a number of ways. 

We will prove some properties of the shape function in specific cases. We also study $\mathbb{T}^3\times\mathbb{R}$, the direct product of the untwisted three-torus and the real line. Identifying the latter as Euclidean time, this is arguably a slightly more realistic model of our actual universe. Again we will see that once the minimal lengths of the non-contractible loops are sufficiently different from each other, $\Sh$ turns negative, with the consequence that $\mathbb{T}^3$ cannot be arbitrarily small (since this would cause the RG flow to end prematurely). 

Then we allow in both examples the torus to become twisted. As in the untwisted case, the shape function for $\mathbb{T}^4$ is invariant under the map to the dual torus, implying a four-dimensional analogue of the moduli space familiar from one loop computations in String Theory \cite{Shapiro:1972ph,Polchinski:1985zf}. However this duality symmetry does not hold for $\mathbb{T}^3$. Although we continue to find that once the non-contractible lengths are sufficiently different, $\Sh$ turns negative, we see that toric universes prefer to twist, in the sense that this increases $\Sh$. For examples with moderately different non-contractible lengths, we will see that a maximal twist is preferred. But on increasing the inhomogeneity, new maxima and minima appear with an apparent arithmetical regularity,  sufficient eventually to cause the domain over which $\Sh>0$ to break into disconnected regions. Therefore in these cases, sufficiently small universes belong to 
disconnected sets labelled by different ranges of the twist parameters.

Through the survey of such examples, we reinforce the conclusion that introducing inhomogeneity already at the $O(1)$ level is sufficient to turn $\Sh$ negative and thus to trigger the potentially important phenomenological effects described above. Clearly however we have only scratched the surface of both the mathematics and the physics in even in these simplest examples.

In sec. \ref{sec:conclusions}, we further discuss the significance of these findings and draw our conclusions.

\section{Review}
\label{sec:review}

In this section we give a review of minimum material from ref. \cite{Morris:2018mhd} that is needed to set the stage for what follows. Wilsonian RG properties are key to understanding whether a quantum field theory has a continuum limit \cite{Wilson:1973,Morris:1998}. As already mentioned, in order to study these properties one has to work with a Euclidean signature metric. Then the Einstein-Hilbert action\footnote{$\kappa^2 = 32\pi G$, $R_{\mu\nu}=R^\alpha_{\ \mu\alpha\nu}$, and $[\nabla_\mu,\nabla_\nu]v^\lambda = R_{\mu\nu\phantom{\lambda}\sigma}^{\phantom{\mu\nu}\lambda}v^\sigma$.}
\be 
\label{EH}
S_{EH} =  \int\!\! d^4x \, \cL_{EH}\,,\qquad \cL_{EH} = -2\sqrt{g} R/\kappa^2\,,
\ee
is unbounded from below, 
so that the Euclidean partition function 
\be 
\label{Z}
\mathcal{Z} = \int\!\! \mathcal{D}g_{\mu\nu}\ {\rm e}^{-S_{EH}}
\ee
will fail to converge. In fact it is unbounded from both above \emph{and} below, as is immediately clear from the fact that the manifold can have arbitrarily large positive or negative curvature. However it is the unboundedly large positive scalar curvature that causes the partition function to be (more than usually) ill-defined.

The Wilsonian RG nevertheless offers a way to make sense of this partition function. At the Gaussian fixed point (necessarily therefore on $\mathbb{R}^4$) and in a particular Feynman -- De Donder gauge, the problem is isolated in the wrong sign for the kinetic term for the trace of fluctuation (the so-called conformal factor or dilaton component):
\be 
\label{Gaussian}
\cL^{\rm kinetic}_{EH} = \frac12 \left(\partial_\lambda \h_{\mu\nu}\right)^2 -\frac12 \left(\partial_\lambda\ph \right)^2\,,
\ee
where we have expressed the metric to first order in $\kappa$ as 
\be 
\label{param-perturbative}
g_{\mu\nu} = \delta_{\mu\nu} \left( 1+\frac{\kappa}{2}\,\ph \right) +\kappa \,\h_{\mu\nu}\,,
\ee
$\h_{\mu\nu}$ is traceless and $\ph$ is the overall local rescaling of the metric. The authors of ref. \cite{Gibbons:1978ac} proposed to fix the problem by continuing the conformal factor functional integral along the imaginary axis: $\ph\mapsto i\ph$. Instead, we will use the Wilsonian RG itself to understand what to do with this ``conformal factor instability''.  

\TRM{We pause to explain why the Wilsonian RG offers a new route, see also ref. \cite{Morris:2018mhd}. The essence of the idea is to recast the definition of the functional integral in differential form. To do this we change the kinetic terms by incorporating multiplicatively a cutoff profile $C^\Lambda(p)$ as in \eqref{total-} and \eqref{DeltaUV}, whose properties are specified below. If the conformal factor had the right sign kinetic term, the effect of the cutoff profile would be to suppress high momentum modes in the functional integral. From this modification we can derive the Wilsonian RG flow equation. We can also reverse the derivation to get the partition function from the flow equation, so the two can be seen as equivalent ways to define the quantum field theory.  
For the wrong sign kinetic term, the cutoff profile  enhances the instability at large momentum, and so only aggravates the problem with the functional integral. Nevertheless, as reviewed below,  we can formally write down the corresponding Wilsonian RG flow equation. Furthermore, the cutoff function will still play its essential practical r\^ole in regularising momentum integrals in the ultraviolet, provided it has the properties specified below. Since the partition function no longer makes sense directly, we take the view that the theory should instead be defined by the flow equation. Indeed since the definition of the theory is then achieved via its RG properties, which are key to understanding a continuum limit, it is  arguably better motivated than analytically continuing the conformal factor along the imaginary axis. 
Unlike the functional integral, we are not immediately faced with difficulties in defining the quantum field theory via this differential equation route. In particular we are assured a quasi-local solution for the effective action (one that can be expanded in derivatives) provided that $C^\Lambda(p)$ has the properties below.\footnote{\TRM{And of course provided that we choose the fields also to be smooth, unlike those that dominate in any functional integral.}} Nevertheless we will see that the flow equation does imply some profound consequences for the theory.}

After integrating out high momentum modes, we can rewrite a continuum partition function exactly in terms of a Wilsonian effective action \cite{Wilson:1973,Morris:1993}. Discarding the traceless component from now on (it will not be needed in this paper), \TRM{\emph{formally}} we can write in this case:
\be 
\label{total-}
S^{\mathrm{tot},\Lambda}[\ph] = S^\Lambda[\ph] - \frac{1}{2}\ph\cdot (\Delta^{\Lambda})^{\!-1}\!\!\cdot \ph\,,
\ee
where $S^\Lambda$ contains the effective interactions and the minus sign in front of the kinetic term signals the instability. The massless propagator
\be 
\label{DeltaUV}
\Delta^\Lambda(p) := \frac{C^\Lambda(p)}{p^2}
\ee
is  regularised by some smooth ultraviolet cutoff profile $C^\Lambda(p)\equiv C(p^2/\Lambda^2)$.  Qualitatively, for $|p|<\Lambda$, $C^\Lambda(p)\approx1$ and mostly leaves the modes unaffected, while for $|p|>\Lambda$ its r\^ole is to suppress modes \TRM{(but see the comments above)}.
We require that $C(p^2/\Lambda^2)$ is a monotonically decreasing function of its argument, that $C^\Lambda(p) \to 1$ for $|p|/\Lambda\to0$, and for $|p|/\Lambda\to\infty$, $C^\Lambda(p) \to0$  sufficiently fast to ensure that all momentum integrals are regulated in the ultraviolet.

After discarding a field independent part, the interactions satisfy the Wilson/Polchinski flow equation \cite{Polchinski:1983gv,Morris:1993}
	\begin{equation}
	\label{pol-}
	\frac{\partial}{\partial\Lambda} S^\Lambda[\ph] ={-}
	\frac{1}{2}\,\frac{\delta S^\Lambda}{\delta\ph}\cdot \frac{\partial\Delta^\Lambda}{\partial\Lambda}\cdot			\frac{\delta S^\Lambda}{\delta\ph}+\frac{1}{2}\,\text{tr}\bigg[\frac{\partial\Delta^\Lambda}{\partial\Lambda}\cdot \frac{\delta^{2}S^\Lambda}			{\delta\ph\delta\ph}\bigg]\,.
	\end{equation}
The first term on the right hand side encodes the tree level corrections, while the second term encodes the quantum corrections. The wrong sign kinetic term leads to an overall sign on the right hand side compared to the usual situation. \TRM{As reviewed above,} at first sight it now looks harmless, but as we will see shortly it has profound consequences.

The Gaussian fixed point is the trivial solution $S^\Lambda[\ph]=0$. We will only need the expression for the eigenoperators, which we can obtain by linearising around the fixed point:
\begin{equation}
	\label{d-pol-}
	\frac{\partial}{\partial \Lambda}\,\delta S^\Lambda[\ph]=\frac{1}{2}\,\text{tr}\bigg[\frac{\partial\Delta^\Lambda}{\partial \Lambda}\cdot \frac{\delta^{2} }{\delta\ph\delta\ph}\bigg] \delta S^\Lambda[\vp]\,.
	\end{equation}
As part of a continuum Wilsonian effective action defined via a smooth cutoff function, such operators have a derivative expansion. 
We saw in ref. \cite{Morris:2018mhd} that the form of the general eigenoperator can then be deduced from the non-derivative (pure potential) eigenoperators. \TRM{We therefore concentrate on such non-derivative operators. We are not of course advocating that these pure potential interactions \textit{per se} provide us with a complete theory of gravity. However their form and RG behaviour is also found for all the interactions that will be involved the theory of gravity \cite{Morris:2018mhd,Morris:2018upm,Morris:2018axr}, as we further explain at the end of this review section.} Writing 
\be 
\label{linearised-k}
\delta S^\Lambda = -\epsilon\! \int\!\!d^4x\,V\!\left(\ph(x),\Lambda\right)\,,
\ee
one obtains
\be 
\label{flow-V}
\partial_t V(\vp,t) = -\Omega_\Lambda\, \partial_\vp^2 \,
V(\vp,t)\,,
\ee 
where $t= \ln(\mu/\Lambda)$ is the RG `time', increasing towards the infrared ($\mu$ some arbitrary energy scale), and we have written the regularised tadpole integral as
\be 
\label{Omega}
\Omega_\Lambda =  |\langle \ph(x) \ph(x) \rangle | = \int\!\frac{d^4p}{(2\pi)^4}\, \Delta^\Lambda(p)\,.
\ee

The RG flow \eqref{flow-V} is reminiscent of a heat diffusion equation.\footnote{A trivial change of variables maps it exactly to the heat equation \cite{Morris:2018mhd}.} The `wrong sign' on the right hand side reverses the direction in which the solutions are well posed, meaning that now only flows to the ultraviolet (UV) exist in general \cite{Bonanno:2012dg,Dietz:2016gzg}. For a generic bare potential at $\Lambda=\Lz$, flow to the IR will end prematurely in a singular effective potential at some critical scale. As we will see, it is this effect that lies at the heart of the restrictions on inhomogeneity.

In fact as shown in ref. \cite{Morris:2018mhd}, without further input the situation is much worse than this. The eigenspectrum degenerates: it includes continuous components, and notions of completeness that exist for normal quantum field theories are now lost \cite{Dietz:2016gzg}. However they are recovered if we now impose a quantisation condition. We insist that the bare potential is square integrable under the following measure:
\be 
\label{QC}
\int^\infty_{-\infty}\!\!\!\! d\vp\,\, {V^{}}^2\!(\vp,\Lambda)\, \exp\left(\frac{\vp^2}{2\Omega_\Lambda}\right)<\infty\,,
\ee
where at the bare level we set $\Lambda=\Lz$. As shown in ref. \cite{Morris:2018mhd}, the RG evolved potential then satisfies this condition for all $\Lambda>\Lz$, and the interactions form a Hilbert space spanned by the eigenoperators ($n$ a non-negative integer)
\be
\label{physical-dnL}
\dd{\Lambda}{n} := \frac{\partial^n}{\partial\vp^n}\, \dd{\Lambda}{0}\,, \qquad{\rm where}\qquad \dd{\Lambda}0 := \frac{1}{\sqrt{2\pi\Omega_\Lambda}}\,\exp\left(-\frac{\vp^2}{2\Omega_\Lambda}\right)\,,
\ee
whose scaling dimensions are their engineering dimensions \ie $-1-n$. They thus form an infinite tower of relevant operators. Other interactions involving \eg space-time derivatives and $\h_{\mu\nu}$, are built with one of these operators as an overall factor and again have scaling dimensions set by their engineering dimensions. They thus also contain infinitely many relevant interactions.  It is this property that holds promise for finding a renormalizable theory of quantum gravity. It should contrasted with the usual approach in quantum gravity, where all the interactions in the Einstein-Hilbert term are irrelevant (and furthermore do not satisfy completeness relations unless we do make the change $\ph\mapsto i\ph$ \cite{Dietz:2016gzg}).

Subject to \eqref{QC}, the general solution of \eqref{flow-V} is thus obtained by taking 
\be 
\label{expand-V}
V(\vp,\Lambda)= \sum_{n=0}^\infty g_n \,\dd\Lambda{n}\,,
\ee
for $\Lambda$-independent couplings $g_n$ with mass dimension $[g_n] = 5+n$. If the couplings are chosen so that the flow exists into the far IR, one can extract the physical potential (we mean the potential in the Legendre effective action) from
\be 
V_\p(\vp) = \lim_{\Lambda\to0} V(\vp,\Lambda)\,.
\ee
Such solutions are characterised by a dynamically generated \emph{amplitude suppression scale} $\Lambda_\p\ge0$ such that
\be 
\label{Vplargephi}
V_\p(\vp)\sim {\rm e}^{-\vp^2/\Lambda_\p^2}
\ee
for large $\vp$ \cite{Morris:2018mhd}. Up to a non-universal constant, this scale also marks the point where the IR evolved potential leaves the Hilbert space \eqref{QC} (because the integral no longer converges for large $\vp$). The potential at any $\Lambda>0$ can be deduced from the physical potential. In fact
\be 
\label{fourier-sol}
V(\vp,\Lambda) = \int^\infty_{-\infty}\!\frac{d\vpi}{2\pi}\, \fV_\p(\vpi)\, {\rm e}^{
-\frac{\vpi^2}{2}\Omega_\Lambda+i\vpi\vp} \,, 
\ee
where the integral is over the dilaton's conjugate momentum $\vpi$, and $\fV_\p$ is the Fourier transform of $V_\p$, as follows from setting $\Lambda=0$. 

\TRM{As we already noted, other interactions in the full theory of quantum gravity must be built with one of the $\dd\Lambda{n}$ as a factor. Including all the relevant couplings, the general form of the interactions is a sum over terms \cite{Morris:2018mhd,Morris:2018axr}
\be 
f^\sigma_\Lambda(\vp)\,\sigma(\partial_\alpha,h_{\beta\gamma},\partial_\delta\vp) + \cdots\,,
\ee
where $\sigma$ is a Lorentz invariant monomial involving some or all of the components indicated (and thus $h_{\mu\nu}$ can appear here differentiated or undifferentiated or not at all),}\footnote{\TRM{General interactions also involve ghosts and extra BRST structure \cite{Morris:2018axr}.}}
\TRM{the ellipses indicate certain tadpole corrections, and
\be 
\label{coefff}
f^\sigma_\Lambda(\vp) = \sum^\infty_{n=n_\sigma} g^\sigma_n \dd\Lambda{n}
\ee
is a \emph{coefficient function} whose properties at linearised level are thus the same as the properties of   the potential operators \eqref{expand-V}, see also \cite{Morris:2018axr}. Therefore we only need to study the behaviour of these latter operators.}
In ref. \cite{Morris:2018mhd}, perturbation theory beyond this linearised level was also considered. However the effects we are concentrating on are already present even at vanishing couplings. \TRM{Therefore} from now on we concentrate on cases where at the linearised level, \TRM{a} physical potential exists on $\mathbb{R}^4$, and ask what form it takes on other manifolds.

\subsection{RG evolution on a manifold}
\label{sec:manifold}

On a manifold $\mathcal{M}$ that is not $\mathbb{R}^4$, the bare operators are still the same, because these operators are defined at $\Lz$, the UV scale that is eventually diverging, corresponding to vanishing distances where the spacetime is indistinguishable from $\mathbb{R}^4$. However the quantum corrections are modified at long distances by the spacetime geometry. To compute this we pull the $\Omega_\Lambda$ term out of the integral in \eqref{fourier-sol} to get
\be 
\label{bare-delta-Omega}
\dd{\Lz}n = \exp\left(\frac12\Omega_\Lz \frac{\partial^2}{\partial\vp^2}\right) \dd{}n\,,
\ee
where $\dd{}n$ is the $n^\mathrm{th}$ derivative of a Dirac delta function: the physical (\ie $\Lambda\to0$) limit of the eigenoperators \eqref{physical-dnL}. Starting with the bare operator, and solving \eqref{d-pol-} on $\mathcal{M}$ down to $\Lambda=k$ gives
\be 
\label{tadpoles-evolution}
\int_x \dd{k}n = \exp\left(-\frac{1}{2}\,\text{tr}\left[\Delta^\Lz_k\cdot \frac{\delta^{2} }{\delta\ph\delta\ph}\right]\right) \int_x \dd{\Lz}n\,,
\ee 
where the explicit space-time integrals and those implied in the space-time trace are now accompanied by $\sqrt{g}$ where $g_{\mu\nu}$ is the (background) metric on $\mathcal{M}$. Here $\Delta^\Lz_k$ is to be the curved space version of the flat space propagator 
\be 
\label{DeltaUVIR}
\frac{C^\Lz_k(p)}{p^2}\,,
\ee
where
\be
\label{CLk}
C^\Lz_k(p) = C^\Lz(p) - C^k(p)
\ee
regulates both the UV and the IR. 

We do not yet know the full theory of quantum gravity incorporating the effects we are describing. Therefore we cannot know for sure how to interpret \eqref{tadpoles-evolution} when the metric is non-trivial. This is one of the main motivations for working later with examples where we can set $g_{\mu\nu}=\delta_{\mu\nu}$. We comment further in the Conclusions. Nevertheless we see that by combining \eqref{tadpoles-evolution} and \eqref{bare-delta-Omega}, the evolved operators can be written as
\be 
\label{delta-M-Omega}
\dd{\!k,\Lz}n = \exp\left(\frac12\,\Omega_{k,\Lz}(x)\, \frac{\partial^2}{\partial\vp^2}\right) \dd{}n\,,
\ee
where
\be 
\label{Omega-M-kL}
\Omega_{k,\Lz}(x) = 
|\langle \ph(x) \ph(x) \rangle |_{\mathbb{R}^4}-|\langle \ph(x) \ph(x) \rangle |_{\mathcal{M}}\,.
\ee 
The first term on the right hand side is just \eqref{Omega} at $\Lambda=\Lz$, while the second term is from propagation on the manifold $\M$ regulated by $C^\Lz_k$. On $\mathbb{R}^4$, this second term is just $\Omega_\Lz-\Omega_k$ and thus $\Omega_{k,\Lz}=\Omega_k$. However on $\M$, $\Omega_{k,\Lz}$ no longer evolves self-similarly but instead picks up ``finite size'' corrections once $k\sim1/L$, where $L$ is some length scale inherent to the manifold. Taking the limits we get the physical $\Omega$:
\be 
\label{Omega-p}
\Omega_\p(x) := \lim_{\Lz\to\infty\atop k\to0} \Omega_{k,\Lz}(x)\,,
\ee
which we can expect (and will verify in particular examples) is a finite universal function of the geometry. Putting $\Omega$ in \eqref{delta-M-Omega} back inside the $\vpi$ integral, we thus get the physical eigenoperators $\dd{\p}{n}$, which therefore have the same form as \eqref{physical-dnL}:
\be
\label{physical-p}
\dd\p{n} = \frac{\partial^n}{\partial\vp^n}\, \dd\p0\,, \qquad{\rm where}\qquad \dd\p0 = \frac{1}{\sqrt{2\pi\Omega_\p}}\,\exp\left(-\frac{\vp^2}{2\Omega_\p}\right)\,.
\ee
Evidently, $\Omega_\p=0$ if the manifold is $\mathbb{R}^4$, and we return to  $\dd{\p}{n}=\dd{}n$. Otherwise, by dimensions 
\be 
\label{shape}
\Omega_\p(x) = \frac{\Sh(x)}{4\pi L^2}\,,
\ee
where $\Sh$ is a (universal) dimensionless `shape' function that can thus only depend on dimensionless characterisations of $\M$. 
Providing $\Sh(x)>0$, $\Omega_\p$ acts to suppress large amplitudes $\vp>1/L$. 
However $\Sh$ can also be negative. 
When this happens, the operators $\dd{k}n$ themselves cease to exist below some critical IR cutoff. However the potential is given by \eqref{fourier-sol} with $\Omega_\Lambda$ replaced by $\Omega_{k,\Lz}(x)$, and can continue to survive below this scale. Fourier transforming \eqref{Vplargephi}, one shows that the flow exists down to $k\to0$, and thus the quantum field theory itself exists, if and only if
\be 
\label{lower-bound}
\Sh(x) > -2\pi L^2\Lambda^2_\p\qquad \forall x\in \M\,.
\ee
Equality would lead to a distributional physical potential at those points, while if the inequality is satisfied, the physical potential has large $\vp$ behaviour:
\be 
\label{compact-large-field}
V_\p\left(\vp(x),x\right) \sim \exp\left(-\frac{\vp^2(x)}{\Lambda_\p^2+2\Omega_\p(x)}\right)\,.
\ee

Let $\Sh_{\rm min}$ be the infimum value of $\Sh$, $\forall x\in\M$. If $\M$ is such that $\Sh_{\rm min}$  is negative, we see from \eqref{lower-bound} that the manifold must have a minimum size 
\be 
\label{Lmin}
L> L_{\rm min}= \frac1{\Lambda_\p}\sqrt{\frac{-\Sh_{\rm min}}{2\pi}}\,.
\ee
We see that the amplitude suppression scale, $\Lambda_\p$, also sets the minimum size of such manifolds. 
Let $\Sh_{\rm max}>0$ be the supremum value for $\Sh_{\rm min}$ over a suitable set of manifolds $\M$ with the same topology. We will see in examples that larger $\Sh$ tends to be associated with more symmetric manifolds, and $\Sh_{\rm max}$ is a number of $O(1)$. For a given manifold $\M$ in the set, we interpret the quantity $\I_\M= \Sh_{\rm max} -  \Sh_{\rm min} >0$ as a measure of its inhomogeneity. It is universal, in the sense of being independent of the details of regularisation. Evidently according to this definition, manifolds with smaller $\Sh_{\rm min}$ are more inhomogeneous. Manifolds that are sufficiently inhomogeneous have negative $\Sh_{\rm min}$. Rephrasing \eqref{Lmin}, the inhomogeneity is bounded above depending on the size of the universe:
\be 
\label{inhomogeneity}
\I_\M < \Sh_{\rm max} + 2\pi L^2\Lambda_\p^2\,.
\ee
In the rest of the paper we will explore the properties of $\Sh$ in toy examples. Of course these examples remain very rudimentary compared to the situation in the real universe. Nevertheless we will uncover already a number of novelties, and demonstrate the extent to which $\I_\M$ does indeed conform to intuitive expectations for inhomogeneity.

\subsection{Four-torus}
\label{sec:four-torus}

In ref. \cite{Morris:2018mhd}, $\Omega_\p$ was computed for the four-torus $\M=\mathbb{T}^4$. The result was found to be
\be 
\label{four-torus}
\Omega_\p = \frac{\Sh_4(\ell_\mu)}{4\pi\sqrt{V_4}}\qquad{\rm where}\qquad \Sh_4(\ell_\mu) = 2-s_4(\ell_\mu)-s_4(1/\ell_\mu)\,.
\ee
Here $\ell_\mu = L_\mu/L$, are dimensionless ratios formed from the four minimal lengths $L_\mu$ of the non-contractible loops, and the geometric mean length scale $L = V^{\frac14}_4$, where the four-volume $V_4= \prod^4_{\mu=1} L_\mu$. The function $s_4$ was defined as
\be 
\label{s4}
s_4(\ell_\mu) = \int_0^1 \frac{dt}{t^2} \bigg( \prod_{\mu=1}^4\, \Theta\left({\ell^2_\mu}/{t}\right)-1\bigg)\,,
\ee
where $\Theta$ is the third Jacobi theta function (at Jacobi $\nu=0$, $x>0$):
\be 
\label{Jacobi}
\Theta(x) = \sum_{n=-\infty}^\infty\!\! {\rm e}^{-\pi n^2 x}\,.
\ee

\section{Spatial three-torus}
\label{sec:three}

 In this section we set $\M$ to be slightly more realistic, namely $\mathbb{T}^3\times\mathbb{R}$, where the real line is to be identified with time (on Wick rotating back to Minkowski signature), and the three-torus is to be identified with the spatial submanifold. As for the four-torus \cite{Hasenfratz:1989pk}, this computation can be related to those in the literature discussing finite size effects in lattice quantum field theory \cite{Hayakawa:2008an}. 

Using \eqref{DeltaUVIR} and \eqref{CLk}, the tadpole integral is
\be 
\label{T3}
|\langle\varphi(x)\varphi(x)\rangle|_{\mathcal{M}}=\frac{1}{V_3}\sum_{n\neq 0}\int\frac{dp_4}{2\pi}\frac{C_k^{\Lambda_0}(p)}{p^2}\,,
\ee 
where $n\in\mathbb{Z}^3\backslash\{0\}$, $p=(p_n^i,p_4)$ and $p_n^i={2\pi n^i}/{L_i}$, where $i=1,2,3$ (no sum over $i$), the $L_i$ are the (minimum) lengths on the non-contractible loops on $\mathbb{T}^3$ and $V_3=\prod_{i=1}^3L_i$ is the volume of $\mathbb{T}^3$. Similar (somewhat) to the $\mathbb{T}^4$ case \cite{Morris:2018mhd}, and following ref. \cite{Hayakawa:2008an}, we remove the $n=0$ zero mode which would otherwise render \eqref{T3} IR divergent.
This can presumably be justified along similar lines \cite{Hayakawa:2008an}, however again such an argument would require first a better understanding of the full gravity theory. With the $n=0$ mode removed, the sum in \eqref{T3} is manifestly IR finite, so that just as in the $\mathbb{T}^4$ case, the limit $k\to0$ in \eqref{Omega-p} can be safely taken, and $\Omega_\p$ is clearly independent of the choice of IR regularisation.

Now we add and subtract the zero mode so that we can use the Poisson summation formula: 
\begin{align}
|\langle\varphi(x)\varphi(x)\rangle|_\mathcal{M}&=\frac{1}{V_3}\sum_n\int\frac{dp_4}{2\pi}\frac{C_k^{\Lambda_0}(p)}{p^2}-\frac{1}{V_3}\int\frac{dp_4}{2\pi}\frac{C_k^{\Lambda_0}(p_4)}{p_4^2}\,,\\
&=\int\frac{d^4p}{(2\pi)^4}\frac{C_k^{\Lambda_0}(p)}{p^2}\sum_n {\rm e}^{i\vec{l}_n\cdot\vec{p}}-\frac{1}{V_3}\int\frac{dp_4}{2\pi}\frac{C_k^{\Lambda_0}(p_4)}{p_4^2}\,,
\end{align}
where the $l_n^i = n^i L_i$ (no sum over $i$) are the windings round the torus, and $n\in\mathbb{Z}^3$.
Just as in the $\mathbb{T}^4$ case, the $n=0$ part of the first term yields $\Omega_{\Lambda_0}-\Omega_k$, and so from \eqref{Omega-M-kL} we get:
\be 
\label{OmegaWinders}
\Omega_{k,\Lambda_0}
=\Omega_k+\frac{1}{V_3}\int\frac{dp_4}{2\pi}\frac{C_k^{\Lambda_0}(p_4)}{p_4^2}-\int\frac{d^4p}{(2\pi)^4}\frac{C_k^{\Lambda_0}(p)}{p^2}\sum_{n\neq 0}{\rm e}^{i\vec{l}_n\cdot\vec{p}}\,.
\ee
As for $\mathbb{T}^4$, UV-finiteness of this expression is now also clear (in particular, of each term), so we are free to use whatever (smooth) regularisation we like, and to take the $\Lambda_0\rightarrow\infty$ whenever we like. From now on, the left hand side will be replaced by $\Omega_\p$ and the implicit limit on the right hand side of $\Lambda_0\rightarrow\infty$ and $k\rightarrow 0$, will be understood. 
Note that the first term, $\Omega_k$, is zero in the IR.

We take the regulator to be 
\be 
\label{cutoff}
C^\Lambda(p)={\rm e}^{-{p^2}/{\Lambda^2}}
\ee 
(as in ref. \cite{Morris:2018mhd}). Using a Schwinger parameter, we rewrite the second term in \eqref{OmegaWinders} as 
\be 
\label{divIR}
\frac{1}{V_3}\int\frac{dp_4}{2\pi}\frac{1}{p_4^2}\left(1-{\rm e}^{-\frac{p_4^2}{k^2}}\right) 
=\frac{1}{2\pi V_3}\int \!dp_4\int_0^{\frac{1}{k^2}}\!\!\!\!d\alpha\,{\rm e}^{-\alpha p_4^2} 
=\frac{1}{2\pi V_3}\int_0^{\frac{1}{k^2}}\!\!\!\!d\alpha\,\sqrt{\frac{\pi}{\alpha}} 
=\frac{1}{k\sqrt{\pi}V_3}\,,
\ee
which is IR divergent. This divergence must therefore be cancelled by the third term.
Treating the third term similarly:
\be 
\int\!\!\frac{d^4p}{(2\pi)^4}\frac{1}{p^2}\left(1-{\rm e}^{-\frac{p^2}{k^2}}\right)\sum_{n\neq 0} {\rm e}^{i\vec{l}_n\cdot\vec{p}} =\int\!\!\frac{d^4p}{(2\pi)^4}\int_0^{\frac{1}{k^2}}\!\!\!\!d\alpha\,{\rm e}^{-\alpha p^2}\sum_{n\neq 0}{\rm e}^{i\vec{l}_n\cdot\vec{p}}
=\frac{1}{4\pi L^2}\int_0^{\frac{4\pi}{k^2L^2}}\frac{dt}{t^2}\sum_{n\neq 0}\prod_{i=1}^3{\rm e}^{-\frac{\pi L_i^2n_i^2}{L^2t}}\,,
\ee
where we integrated over $p$ and used the change of variables $\alpha=\frac{L^2t}{4\pi}$, where $L=V_3^{\frac{1}{3}}$. We can add and subtract the $n=0$ part in order to write this in terms of the third Jacobi theta function \eqref{Jacobi}:
\be 
\frac{1}{4\pi L^2}\int_0^{\frac{4\pi}{k^2L^2}}\frac{dt}{t^2}\left(\prod_{i=1}^3\Theta\left(\ell^2_i/t\right)-1\right)\,,
\ee 
where we have written $\ell_i={L_i}/{L}$. 
In order to deal with this integral, we first define
\be 
\label{s3}
s_3(\ell_i)=\int_0^1\frac{dt}{t^2}\left(\prod_{i=1}^3\Theta\left({\ell_i^2}/{t}\right)-1\right)
\ee 
so that we can write
\be 
\int_0^{\frac{4\pi}{k^2L^2}}\frac{dt}{t^2}\left(\prod_{i=1}^3\Theta\left(\ell^2_i/t\right)-1\right)=s_3(\ell_i)+\int_1^{\frac{4\pi}{k^2L^2}}\frac{dt}{t^2}\left(\prod_{i=1}^3\Theta\left({\ell_i^2}/{t}\right)-1\right)\,.
\ee 
Taking $t\mapsto {1}/{t}$ in the remaining integral and using the identity $\Theta(x)=\frac{1}{\sqrt{x}}\Theta\left(\frac{1}{x}\right)$, it becomes
\be 
\label{s3-twiddle}
\int_{\frac{k^2L^2}{4\pi}}^1\!\!dt\left(\prod_{i=1}^3\left(\frac{1}{\sqrt{\ell_i^2t}}\Theta\left(\frac{1}{\ell_i^2t}\right)\right)-1\right)   
=\tilde{s}_3\left({1}/{\ell_i}\right)+\int_{\frac{k^2L^2}{4\pi}}^1dt\left(t^{-\frac{3}{2}}-1\right)
\ee
where, taking the $k\rightarrow 0$ limit, we have defined a new function
\be 
\tilde{s}_3(\ell_i)=\int_0^1\frac{dt}{t^{\frac{3}{2}}}\left(\prod_{i=1}^3\Theta\left({\ell_i^2}/{t}\right)-1\right)\,.
\ee 
Note that the last term in \eqref{s3-twiddle} does indeed cancel the IR divergent \eqref{divIR}, leaving behind a finite part. Putting all bits back together gives the final answer
\be 
\label{three-torus}
\Omega_\p= \frac{\Sh_3(\ell_i)}{4\pi V_3^{\frac{2}{3}}}\quad\text{where}\quad \Sh_3(\ell_i) = 3-s_3(\ell_i)-\tilde{s}_3\left({1}/{\ell_i}\right)\,.
\ee 
This should be compared to the $\mathbb{T}^4$ result \eqref{four-torus}, \eqref{s4}.
As must be, $\Omega_\p$ is a function purely of the geometry of the manifold. Specifically, for both $\mathbb{T}^4$ and $\mathbb{T}^3\times\mathbb{R}$, it only depends on the lengths of the fundamental loops. We see that in both cases there is a symmetry under interchange of the lengths, as expected from the symmetries of the torii. However, the ``T-duality-like" symmetry $L_\mu\mapsto L^2/L_\mu$ that was present for the four-torus \cite{Morris:2018mhd}, is no longer present  since $\tilde{s}_3\neq s_3$. 

To get some intuition, we input some numerical values:
\begin{itemize}
\item $\ell_1=\ell_2=\ell_3=1$: $\Sh_3=2.8373$;
\item $\ell_1=1$, $\ell_2=2$, $\ell_3=\frac{1}{2}$: $\Sh_3=0.8538$;
\item $\ell_1=1$, $\ell_2=3$, $\ell_3=\frac{1}{3}$: $\Sh_3=-4.2936$;
\item $\ell_1=\ell_2=2$, $\ell_3=\frac{1}{4}$: $\Sh_3=-8.95463$;
\item $\ell_1=\ell_2=3$, $\ell_3=\frac{1}{9}$: $\Sh_3=-73.1222$;
\item $\ell_1=2$, $\ell_2=3$, $\ell_3=\frac{1}{6}$: $\Sh_3=-28.4098$;
\item $\ell_1=\frac{1}{2}$, $\ell_2=\frac{1}{3}$, $\ell_3=6$: $\Sh_3=-15.7999$.
\end{itemize}
We thus see that $\Sh_3$ is positive for the perfectly symmetric case, decreases as the space becomes less isotropic, and that it can be negative without particularly extreme anisotropy, broadly mirroring the $\mathbb{T}^4$ case. Comparing the final two examples also gives an explicit demonstration that the $\ell_i\mapsto {1}/{\ell_i}$ symmetry no longer holds.

This $\mathbb{T}^3\times\mathbb{R}$ case must satisfy the general constraint \eqref{lower-bound}, so that when $\Sh_3<0$ we have a minimum size for $\mathbb{T}^3$ given by \eqref{Lmin}. Alternatively in this case we can write
\be 
V_3>\left(-\frac{\Sh_3}{2\pi\Lambda_p^2}\right)^{\frac{3}{2}}\,.
\ee 
This again has the phenomenological significance outlined in the Introduction, for example universes of small spatial extent are thus constrained to be strongly symmetric, whereas larger ones are permitted to have more anisotropies.

%
%

\section{The three-torus versus a limit of the four-torus}

Since we can get from $\mathbb{T}^4$ to $\mathbb{T}^3\times\mathbb{R}$ by sending one of the lengths, say $L_4$, to infinity, it might be expected that the two results for $\Omega_\p$ are similarly related. In fact this is not the case if we send $k\to0$ and then send $L_4\to\infty$, which is the physically motivated order in which we should take these IR limits (for example as might happen as a result stretching of $L_4$ due to dynamics). The result does depend on the order. If we keep $k$ fixed, we can compare the results but then they differ by $k\to0$ divergent quantities, as we show below. 

It is also the case that the zero mode subtraction is treated differently in both cases. 
If we treated the four-torus with the same zero-mode subtraction, we would write
\be 
|\langle\varphi(x)\varphi(x)\rangle|_{\mathcal{M}}=\frac{1}{V_4}\sum_{\vec{n}\neq 0}\sum_{n_4}\frac{C_k^{\Lambda_0}(p_n)}{p_n^2}\,,
\ee
where $p^\mu_n = 2\pi n_\mu/L_\mu$ (no summation over $\mu$). Therefore the difference between this and
\be 
\label{four-torus-tadpole}
|\langle \ph(x) \ph(x) \rangle |_{\mathcal{M}} = \frac1{V_4} \sum_{n\ne0} \frac{C^\Lz_k(p_n)}{p^2_n}\,,
\ee
which is the treatment in ref. \cite{Morris:2018mhd},
is just (taking the UV and IR limits):
\be 
\label{zmodecorr}
\frac{L_4^2}{2\pi^2V_4}\sum_{n_4=1}^\infty\frac{1}{n_4^2} = \frac{L_4^2}{12V_4} = \frac{L_4}{12V_3}\,.
\ee

More serious is the exchange of $L_4$ and $k$ limits. Starting with \eqref{four-torus-tadpole} and using
 \eqref{cutoff} to add back the zero mode contribution, $1/(V_4k^2)$, we have, after Poisson resummation,
\be 
\Omega_\p = \frac1{V_4k^2}  -\int \frac{d^4p}{(2\pi)^4}  \frac{C^\Lz_k(p)}{p^2} \sum_{n\ne0} {\rm e}^{il_n\cdot p}\,,
\ee
where again we are taking the UV and IR limits where it is unambiguous, and $l_{n\,\mu} = L_\mu n_\mu$ (not summed over $\mu$) are the windings round $\mathbb{T}^4$. Separating $L_4$ in the second term, we write is as
\be
-\int\!\frac{d^4p}{(2\pi)^4}\int_0^{\frac{1}{k^2}}\!\!d\alpha\,{\rm e}^{-\alpha p^2}\sum_{n\neq 0}{\rm e}^{il_n\cdot p}= -\frac{1}{4\pi L^2}\int_0^{\frac{4\pi}{k^2L^2}}\frac{dt}{t^2}\left(\prod_{\mu=1}^4\Theta\left({\ell_\mu^2}/{t}\right)-1\right)\,,
\ee
where $L=V^{1/3}_3$ is defined from the three-volume and thus also the ratios $\ell_\mu={L_\mu}/{L}$. Splitting the range into two parts as before, we see that for $t\in(0,1)$ the limit $\ell_4\to\infty$ can be safely taken and readily reduces to \eqref{s3}. The $t\in(1,4\pi/k^2L^2)$ part however depends on the order of limits. Keeping $k$ fixed, we can again use $\Theta\left({\ell_4}/{t}\right)\to 1$ as $\ell_4\to\infty$, and thus we find:
\be 
\label{Omegas}
\Omega_\p|_{\mathbb{T}^4} = \Omega_\p|_{\mathbb{T}^3\times\mathbb{R}} + \frac1{k^2L_4V_3}-\frac1{kV_3\sqrt{\pi}}\,,
\ee
where we expose the remaining discrepancies in $L_4$ and $k$. To this we should add \eqref{zmodecorr} depending on the treatment of the zero mode(s), but in any case we see that as claimed the difference is now IR divergent. 

Finally we note that \eqref{Omegas} suggests it is possible to get $\Omega_\p|_{\mathbb{T}^4} \to \Omega_\p|_{\mathbb{T}^3\times\mathbb{R}}$ by taking the two limits together. From \eqref{Omegas} we would set $L_4 = \sqrt{\pi}/k$, up to further terms that might arise from reassessing the limit $\Theta\left({\ell_4}/{t}\right)\to 1$. It is not clear however that such a limiting procedure carries any physical significance.

\section{Analytic results}

We demonstrate that it is possible to establish analytically some properties of the shape function $\Sh_d$. The most symmetric point is a stationary point in each of the cases $d=3,4$, \ie \eqref{four-torus} and \eqref{three-torus}. Indeed both $\Sh_d$ are symmetric under interchange of any of the lengths of the fundamental loops.
Writing $\ell_\mu = {\rm e}^{z_\mu}$, the most general perturbation takes the form $\delta z_\mu = \alpha_\mu\, \epsilon$. At the symmetric point ($\ell_\mu=1$ for $\mu=1,\cdots,d$), the first-order change in $\epsilon$ is thus proportional to $\sum_\mu\alpha_\mu$. But since the $\ell_\mu$ must satisfy the constraint $\prod_\mu \ell_\mu =1$, we have that $\sum_\mu\alpha_\mu=0$, and thus the vanishing of first-order perturbations at the symmetric point.


We will now prove that the symmetric point is the global maximum in the case where just two $\ell\ne1$. Without loss of generality we can take $\ell_1 = \sqrt{\chi}$ and $\ell_2 =1/\sqrt{\chi}$, where $\chi>0$. For both cases \eqref{four-torus} and \eqref{three-torus}, the dependence on the lengths of the fundamental loops is now only in the combination 
\be
\theta(\chi)=\Theta\left(\frac{\chi}{t}\right)\Theta\left(\frac{1}{t\chi }\right)\,.
\ee
Since this is $\chi\mapsto1/\chi$ invariant, we are free to consider only $\chi\in(0,1]$. Using a result of Ramanujan (given \eg in Berndt \cite{MR1117903})
\be 
\ln\Theta(x)=2\sum_{k=1}^\infty\frac{q^{2k-1}}{(2k-1)(1+q^{2k-1})}\,,
\ee
where $q={\rm e}^{-\pi x}$ and recall \eqref{Jacobi}, we have 
\be 
\partial_x \Theta(x) = -2\pi\,\Theta(x)\sum_{k=1}^\infty\frac{1}{\left(1+e^{\pi(2k-1)x}\right)^2}\,,
\ee
and thus
\be 
\partial_\chi\theta(\chi)=\frac{2\pi\theta}{t\chi^2}\sum_{k=1}^\infty\frac{
1+\exp\left(\frac{\pi(2k-1)\chi}{t}\right)+\chi\left(1+\exp\frac{\pi(2k-1)}{\chi t}\right)}{ \left(1+\exp\frac{\pi(2k-1)}{\chi t}\right)^2\left(1+\exp\frac{\pi(2k-1)\chi }{t}\right)^2}f_k(\chi)\,,
\ee
where the sign is determined by
\be 
f_k(\chi) = 1+\exp\left(\frac{\pi(2k-1)\chi}{t}\right)-\chi\left(1+\exp\frac{\pi(2k-1)}{\chi t}\right)\,.
\ee
Noting that $2k-1\ge1$ we have that
\be
f'_k(\chi)\ge\pi\exp\left(\frac{\pi \chi (2k-1)}{t}\right)-1+\exp\left(\frac{\pi(2k-1)}{\chi t}\right)\left(\frac{\pi}{\chi }-1\right)
\ee
which is positive in the range $\chi\in(0,1]$. Since $f_k(1)=0$, we have established that $f_k(\chi)<0$ in this range. This implies that $\theta(\chi)$ has a global minimum at $\chi=1$, which in turn implies the same for $s_3$, $\tilde{s}_3$ and $s_4$, and thus that $\chi=1$ is the global maximum for $\Sh_d$.

\section{Twisted torii}

Although we are restricted at this stage of development, to using a background metric $g_{\mu\nu}=\delta_{\mu\nu}$, we can nevertheless consider a larger set of manifolds with the same topology, namely twisted torii. In this section we begin an investigation of the shape function $\Sh_d$ in these cases and the consequent restriction on the geometry. We will see in particular that the torii actually prefer to twist, in the sense that this increases $\Sh_d$. 

With the background metric set to $g_{\mu\nu}=\delta_{\mu\nu}$, the twisted four-torus is given by the equivalence relation $x^\mu\sim x^\mu+ L v^\mu$, over a lattice $\Latt$ defined by $v^\mu = \sum_{i=1}^4 n^i \ell^\mu_i$, $n^i$ being integers, where the (scaled) primitive vectors $\vec{\ell}_i$ are not all orthogonal. Anticipating the need, we have factored out the length scale $L=V^{1/4}_4$, where $V_4$ is the volume of the torus. Thus the $\vec{\ell}_i$ satisfy  the constraint $\det\left(\vec{\ell}_1,\vec{\ell}_2,\vec{\ell}_3,\vec{\ell}_4\right)=1$, corresponding to unit volume for the unit cell. Considering ${\rm e}^{i\vec{x}\cdot\vec{p}}$, the momentum modes live on the dual lattice, $\vec{p}\in\Latt^*/L$. Having thus factored out the length scale, the dimensionless dual lattice is given by
\be
\Latt^*=\{\vec{u}\in\mathbb{R}^4;\vec{u}\cdot\vec{v}\in 2\pi\mathbb{Z}\ \forall\vec{v}\in\Latt\}\,,
\ee
and also has a unit volume unit cell.
Removing the zero mode as before we thus have
\be 
|\langle\varphi(x)\varphi(x)\rangle|_{\mathcal{M}}=\frac{1}{V_4}\sum_{L\vec{p}\,\in\,\Latt^*\backslash\{\vec{0}\}}\frac{C_k^{\Lambda_0}(p)}{p^2}\,.
\ee
Following the same procedures as before we are led to the lattice theta function,
a well-studied object in the subject of modular forms (see \eg \cite{MR0344216}). It is defined by
\be 
\label{ThetaLatt}
\Theta_{\Latt}(t)=\sum_{v\in\Latt}{\rm e}^{-\pi tv^2}\,,
\ee
and is well-defined for $t>0$. We use the Poisson summation formula, which now takes the form
\be 
\Theta_{\Latt}(t)=t^{-{d}/{2}}\,\Theta_{\Latt^*}\left({1}/{t}\right)
\ee
for a $d$ dimensional lattice. (A factor of inverse volume would also appear if the unit cell had other than unit volume.)
Otherwise following the previous steps, one arrives at
\be 
\label{Omega4twisted}
\Omega_\p=\frac{\Sh_4(\Latt_4)}{4\pi\sqrt{V_4}} \quad\text{where}\quad \Sh_4(\Latt_4)=2-s(\Latt_4)-s(\Latt^*_4)\,,
\ee
defining for a $d$ dimensional lattice 
\be
\label{s}
s(\Latt)=\int_0^1\!\frac{dt}{t^2}\left(\Theta_{\Latt}\left({1}/{t}\right)-1\right)\,.
\ee
Generalising in a similar way to a twisted $\mathbb{T}^3\times\mathbb{R}$, one arrives at
\be 
\Omega_\p= \frac{\Sh_3(\Latt_3)}{4\pi V_3^{\frac{2}{3}}}\quad\text{where}\quad \Sh_3(\Latt_3) = 3-s(\Latt_3)-\tilde{s}\left(\Latt^*_3\right)\,,
\ee
$s$ being given already in \eqref{s}, and 
\be 
\label{stwiddle}
\tilde{s}(\Latt)=\int_0^1\frac{dt}{t^{{3}/{2}}}\left(\Theta_{\Latt}\left({1}/{t}\right)-1\right)\,.
\ee
It is straightforward to see that these formulae coincide with the previous ones in the case that the lattice is untwisted.
Just as for the untwisted cases, it is clear from the sum over momenta that $\Omega_\p$ is safe and universal in the IR, and from the sum over windings that it is safe and universal in the UV.
 The symmetry under permuting the fundamental lengths is now a symmetry under relabelling the primitive vectors $\vec{\ell}_i$. However we also have a full PSL$(d,\mathbb{Z})$ symmetry, the modular group corresponding to the additional freedom to redefine any primitive vector $\vec{\ell}_i$ by adding integer multiples of $\vec{\ell}_{j\ne i}$. Furthermore for $\mathbb{T}^4$ the inversion symmetry noticed in ref. \cite{Morris:2018mhd}, is now generalised to invariance under swopping the torus for its dual torus (\ie $\Latt\leftrightarrow\Latt^*$). 

\subsection{Analytic results}
\label{sec:twistAnalytic}

The simplest example of the additional freedom to deform the torus, is to start with orthogonal vectors $\vec{\ell}_i$  and twist $\vec{\ell}_1\mapsto\vec{\ell}_1+a\vec{\ell}_2$, where without loss of generality the modulus is restricted to the range $a\in[0,1)$, since as we noted $a\in\mathbb{Z}$  defines the same lattice. The dependence of the shape functions $\Sh_d$ on $a$ is then contained in $\Theta_{\Latt_2}(1/t)$ and $\Theta_{\Latt^*_2}(1/t)$ where, as defined in \eqref{ThetaLatt}, the sum over the 2-lattices amounts to sum over integers $n_1,n_2$ where for $v\in\Latt_2$ we have
\be 
v^2 = \ell_1^2n_1^2+\ell_2^2(n_1a+n_2)^2\,,
\ee
and for $v\in\Latt_2^*$ we have
\be 
v^2 = (n_1-an_2)^2/\ell_1^2+n_2^2/\ell_2^2\,.
\ee
In both cases, and thus also for $\Sh_d$, we see that if $\ell_1\gg\ell_2$ the dependence on $a$ is exponentially suppressed. On the contrary, we see that if $\ell_2\gg\ell_1$ the dependence on $a$ is dramatically enhanced. We will see numerical examples of this shortly. 
We also see clearly that when the primitive vectors start off orthogonal, the results are symmetric under $a \mapsto -a$ (after summing over $n_1,n_2$). Since $\Sh_d$ is differentiable, this implies that $\Sh_d$ has a stationary point at zero twist ($a=0$). Combining with the lattice symmetry $a\mapsto a+1$, we see that $\Sh_d$ is also symmetric about maximum twist ($a=1/2$), which is thus another stationary point.


\subsection{Numerical examples}

While Jacobi theta functions are encoded into algebraic computing packages, the same is not true of the lattice theta function. For numerical investigations it is helpful to note that 
\be 
s(\Latt) = \sum_{v\in\Latt\backslash\{0\}}\frac{1}{\pi v^2}\exp-\pi v^2\,,
\ee
as is clear from substituting \eqref{ThetaLatt} into \eqref{s} and performing the integral, and similarly for \eqref{stwiddle}:
\be 
\tilde{s}(\Latt) = \sum_{v\in\Latt\backslash\{0\}}\frac{1}{|v|}\,\erfc\left(\sqrt{\pi v^2}\right)\,.
\ee
By organising the expansion in increasing $|v|$, the shape functions can then be computed with exponentially fast convergence.

\subsubsection{Twisted four-torus}

\begin{figure}[ht]
\begin{center}
$
\begin{array}{cc}
\qquad\qquad M=\begin{pmatrix}
1 & a & 0 & 0\\
0 & 1 & 0 & 0\\
0 & 0 & 1 & 0\\
0 & 0 & 0 & 1
\end{pmatrix}\qquad:\qquad & \hskip-0.2\textwidth
\vcenter{\includegraphics[width=0.37\textwidth]{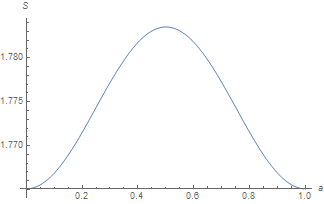}} \\
\end{array}
$
\end{center}
\caption{The shape function $\Sh_4$, twisting $a$ away from orthonormal primitive vectors.}
\label{fig:x1}
\end{figure}


\begin{figure}[ht]
\begin{center}
$
\begin{array}{cc}
\qquad\qquad M=\begin{pmatrix}
1 & 0 & 0 & 0\\
0 & 1 & 0 & 0\\
0 & 0 & 2 & 0\\
0 & 0 & a & \frac{1}{2}
\end{pmatrix}\qquad:\qquad & \hskip-0.2\textwidth
\vcenter{\includegraphics[width=0.37\textwidth]{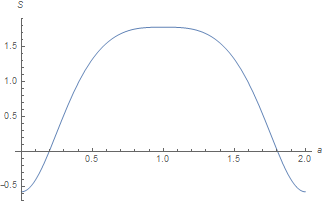}} \\
\end{array}
$
\end{center}
\caption{The shape function $\Sh_4$, twisting by $a$ from a smaller vector towards a larger vector.}
\label{fig:x2}
\end{figure}


We again find that if the lengths of the primitive vectors are sufficiently different the shape function turns negative. However, twisting delays this effect. To show examples, let us write the primitive vectors as rows of a matrix $M$. In fig. \ref{fig:x1} we see as advertised that $\Sh_4$ is symmetric about $a=0,1/2$ and furthermore is maximised at maximum twist (where its value $\Sh_4=1.784$ should be compared to the zero twist result $\Sh_4=1.765$ \cite{Hasenfratz:1989pk,Morris:2018mhd}).
In fig. \ref{fig:x2}, we see how the effect is accentuated when twisting from a smaller vector towards a larger vector, and that again the maximum value is reached at maximal twist (here $a=1$). The numerical values are displayed at the bottom of table \ref{table:twists4}, where we also survey some other examples. Such cases result in larger plateaux around the center, as can clearly be seen above. In part this is simply because in the case of the large vector being added to the smaller one, as $a$ increases up the midpoint, the change in the length of the primitive vector becomes smaller and smaller relative to the length itself. 

\begin{table} 
\begin{center}
\begin{tabular}{|c|c|c|c|}
\hline
Matrix   & $\Sh_{4\, {\rm max}}$ & Range of $\Sh_4$\\ [0.5ex]
\hline\hline
$\begin{pmatrix}
1 & a & 0 & 0\\
0 & 1 & 0 & 0\\
0 & 0 & 2 & 0\\
0 & 0 & 0 & \frac{1}{2}
\end{pmatrix}$ &  $-0.547798$ & $0.0293644$\\
\hline

$\begin{pmatrix}
1 & 0 & a & 0\\
0 & 1 & 0 & 0\\
0 & 0 & 2 & 0\\
0 & 0 & 0 & \frac{1}{2}
\end{pmatrix}$

$\cong$

$\begin{pmatrix}
1 & 0 & 0 & 0\\
0 & 1 & 0 & 0\\
0 & 0 & 2 & 0\\
a & 0 & 0 & \frac{1}{2}
\end{pmatrix}$ 

 & $0.128869$ & $0.706032$\\
\hline
$\begin{pmatrix}
1 & 0 & 0 & a\\
0 & 1 & 0 & 0\\
0 & 0 & 2 & 0\\
0 & 0 & 0 & \frac{1}{2}
\end{pmatrix}$

$\cong$

$\begin{pmatrix}
1 & 0 & 0 & 0\\
0 & 1 & 0 & 0\\
a & 0 & 2 & 0\\
0 & 0 & 0 & \frac{1}{2}
\end{pmatrix}$ 

 & $-0.577106$ & $5.6\times 10^{-5}$\\
\hline

$\begin{pmatrix}
1 & 0 & 0 & 0\\
0 & 1 & 0 & 0\\
0 & 0 & 2 & a\\
0 & 0 & 0 & \frac{1}{2}
\end{pmatrix}$  & $-0.577163$ & $\sim 10^{-10}$\\
\hline

$\begin{pmatrix}
1 & 0 & 0 & 0\\
0 & 1 & 0 & 0\\
0 & 0 & 2 & 0\\
0 & 0 & a & \frac{1}{2}
\end{pmatrix}$ & $1.78352$ & $2.36068$\\
\hline
\end{tabular}
\end{center}
\caption{Examples of twisting. $\Sh_4$ is invariant under $\Latt\mapsto\Latt^*$. Rows with only one matrix, are mapped $a\mapsto-a$ by duality, while those with two matrices display the dual lattice (after some relabelling and reparametrisation). In all cases the maximum value for $\Sh_4$ is reached at maximum twist \ie when $a$ is half the length of the primitive vector we are twisting towards. It should be compared to the untwisted case which for these primitive vectors gives $\Sh_4 = -0.577163$.}
\label{table:twists4}
\end{table}

\begin{figure}[ht]
\begin{center}
$
\begin{array}{ccc}
\includegraphics[width=0.31\textwidth]{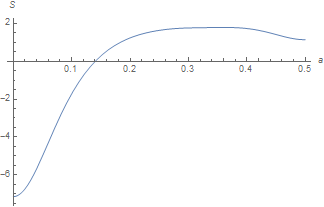} &
\includegraphics[width=0.31\textwidth]{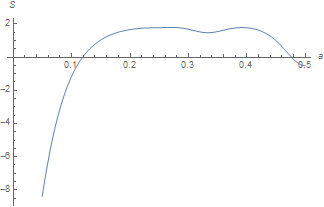} &
\includegraphics[width=0.31\textwidth]{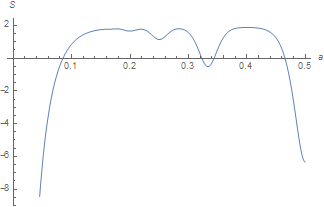} \\
\end{array}
$
\end{center}
\caption{$\Sh_4$ for torii \eqref{Max} with twist $0\le a\le 1/2$, and respectively $x=3,4,6$.}
\label{fig:x346}
\end{figure}

However once the lengths of the primitive vectors are sufficiently different from each other, new effects open up, as we show in fig. \ref{fig:x346}. Here we generalise the previous case and write:
\be 
\label{Max}
M=M_4(a,x)=\begin{pmatrix}
1 & 0 & 0 & 0\\
0 & 1 & 0 & 0\\
0 & 0 & x & 0\\
0 & 0 & ax & \frac{1}{x}
\end{pmatrix}\,. 
\ee
We are following the parametrisation in sec. \ref{sec:twistAnalytic}, so such an $\Sh_4(a,x)$ is symmetric about $a=1/2$ and $a=0$, and periodic with period $a=1$. The symmetry $\Latt\leftrightarrow\Latt^*$ swops $a$ for $-a$.

 For $x=3$ we see that maximum twist has become a local minimum. Increasing to $x=4$, two more local extrema have appeared and significantly, $\Sh$ is now negative at maximum twist. Increasing to $x=6$, we find further regions of $\Sh<0$ opening up. We also note that the minima seem to occur at $\frac{1}{2}$, $\frac{1}{3}$, $\frac{1}{4}$ and $\frac{1}{5}$, moreover the first two correspond to minima in the $x=4$ figure. We have verified this behaviour appears also at non-integer $x$. Thus we are led to conjecture that for given $x$, minima appear at $a=\frac{1}{n}$, for all $n\in\mathbb{N}$ such that $n<x$.

\begin{figure}[ht]
\begin{center}
$
\begin{array}{ccc}
\includegraphics[width=0.35\textwidth]{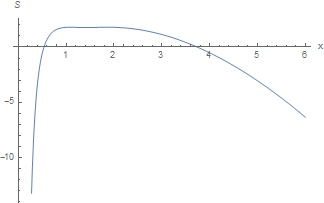} & \qquad &
\includegraphics[width=0.35\textwidth]{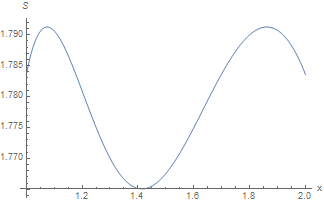} \\
\end{array}
$
\end{center}
\caption{$\Sh_4(\half,x)$ for the torus in \eqref{Max}. The right hand panel is a close-up for $x\in[1,2]$.}
\label{fig:x}
\end{figure}

At fixed $a$ in \eqref{Max}, $x$ scans over inhomogeneity. In fig. \ref{fig:x} we set $a=\half$. We see that for sufficient inhomogeneity in the sense of small enough or large enough $x$, $\Sh_4$ turns increasingly negative, thus broadly following the pattern we found for untwisted torii. However from the right hand panel we see that the maximum $\Sh_4$ occurs at $x\ne1$ for this twisted case. Thus according to the interpretations we gave in sec. \ref{sec:manifold}, with $a$ fixed to $\half$, it is these points that the theory regards as ``most symmetric''. We note that the local minimum appears to be at $x=\sqrt2$. For $\Sh_4(\frac13,x)$ we find similarly that the local minimum appears to be at $x=\sqrt3$. 

\subsubsection{Twisted three-torus}

\begin{figure}[ht]
\begin{center}
$
\begin{array}{ccc}
\includegraphics[width=0.35\textwidth]{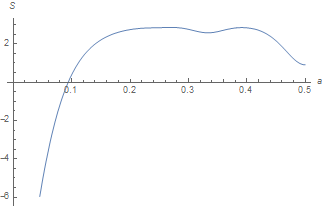} & \qquad &
\includegraphics[width=0.35\textwidth]{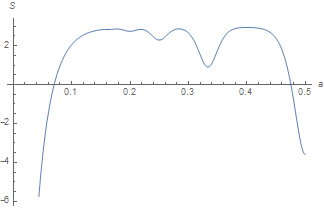} \\
\end{array}
$
\end{center}
\caption{$\Sh_3(a,x)$ for torii \eqref{Max3} with twist $0\le a\le 1/2$, and respectively $x=4,6$.}
\label{fig:T3}
\end{figure}

Numerically we find similar behaviour to twisting $\mathbb{T}^4$, except the effects tend to be more mild. For example, setting
\be 
M=\begin{pmatrix}
1 & 1 & 0\\
0 & 2 & 0\\
0 & 0 & \frac{1}{2}\\
\end{pmatrix}
\ee 
gives $\Sh_3=1.327$, increasing from the untwisted $\Sh_3=0.8538$. And twisting in the direction from a smaller primitive vector, for example 
\be 
M= \begin{pmatrix}
1 & 0 & 0\\
0 & 2 & 0\\
0 & 1 & \frac{1}{2}
\end{pmatrix}
\ee 
increases the shape function still further: $\Sh_3=2.8538$.  Setting 
\be
\label{Max3}
M=M_3(a,x)=\begin{pmatrix}
1 & 0 & 0\\
0 & x & 0\\
0 & ax & \frac{1}{x}
\end{pmatrix}\,,
\ee
we find similar but less marked behaviour to \eqref{Max}, as can be seen in the samples in fig. \ref{fig:T3} (compare the right panels of fig. \ref{fig:x346}).

Although qualitative differences will arise since \eqref{s3-twiddle} is now involved, $\Sh_3(a,x)$ is also periodic under $a\mapsto a+1$, and from sec. \ref{sec:twistAnalytic} we know that it is also symmetric about $a=\half$ and $a=0$. Since swopping $\Latt$ for $\Latt^*$ amounts to exchanging $a$ for $-a$, this last symmetry means that actually both $\Sh_d(a,x)$ depend on $a$ and $x$ only through the corresponding $\Theta_{\Latt_2}(1/t)$ as defined in sec. \ref{sec:twistAnalytic}. Since furthermore the stationary points appear to be in the same place, it must be that $\Theta_{\Latt_2}(1/t)$ itself is stationary at those values.

\section{Conclusions}
\label{sec:conclusions}

If we focus on the conformal mode of the metric then, as shown in refs. \cite{Dietz:2016gzg,Morris:2018mhd} and reviewed in sec. \ref{sec:review}, it has
profoundly altered RG  properties inherited from the fact that in Euclidean signature (where Wilsonian RG makes sense) this mode has the wrong sign kinetic term. (This wrong sign is in turn related to the fact that the Einstein-Hilbert action is unbounded below in Euclidean signature.) In this work we have further developed and explored one aspect of this: the fact that the quantum field theory itself fails to exist once a universal shape function $\Sh$ falls below a lower bound set by the typical length scale $L$ in the manifold and the amplitude suppression scale $\Lambda_\p$, as encapsulated in \eqref{lower-bound}. For finite $\Lambda_\p$, manifolds cannot therefore be arbitrarily small if $\Sh<0$. We confirm in the examples we studied, that this happens once length scales in the manifold are sufficiently different from each other, \ie once the manifold is sufficiently inhomogeneous.  We have seen that this applies not only  to the fundamental loops in $\mathbb{T}^4$ \cite{Morris:2018mhd}, but also to those in $\mathbb{T}^3\times\mathbb{R}$, and to twisted versions of these two cases. 

However as well as this broad inhomogeneity effect, we have also seen that twisting is preferred in the sense that it increases $\Sh$, the amount of twisting preferred itself dependent on the overall inhomogeneity, in an increasingly involved way as the inhomogeneity increases. The length and twisting parameters lie on an intricate moduli space. In particular the shape function for (a twisted) $\mathbb{T}^4$ is invariant under replacement with the dual torus. Overall, it is clear that we have only scratched the surface of the phenomena accessible even in these simplest examples.

Our investigation of shape dependence is essentially a kinematic study of the single component scalar field $\ph$ with wrong sign kinetic term, in the sense that dynamical evolution of the spacetime is so far missing. Presumably a full theory of quantum gravity based on the operator spectrum \eqref{physical-dnL}, would avoid evolving into manifolds where the theory then ceases to exist. A full understanding of these aspects will however have to wait until such a theory is developed.

If one assumes that these effects survive in a fully developed theory of quantum gravity, and that in this theory the background spacetime has a diffeomorphism invariant description using a background metric in the usual way, then it is natural to expect that the conformal factor $\ph$ is conformally coupled to the background metric. We note that this then implies an acceptable (but difficult to study) phenomenology. Thus, since a Friedmann-Lema\^\i tre-Robertson-Walker spacetime is conformally flat, restrictions on the size of the universe will arise only once inhomogeneity is introduced on top of the standard cosmological model.

Finally we recall again that $\Sh$ is closely related to the study of finite size effects in lattice quantum field theory \cite{Hasenfratz:1989pk,Hayakawa:2008an}. To our knowledge finite size effects in twisted lattices have not been addressed, so our results could be adapted to this very different area if ever such cases were considered useful.

\bigskip\bigskip

\section*{Acknowledgments}
MPK acknowledges support via an STFC PhD studentship. TRM acknowledges support from both the Leverhulme Trust and the Royal Society as a Royal Society Leverhulme Trust Senior
Research Fellow, and from
STFC through Consolidated Grants ST/L000296/1 and ST/P000711/1.

\vfill
\newpage 



\bibliographystyle{hunsrt}
\bibliography{references} 

\end{document}